\documentclass[12pt]{article}
\usepackage[utf8]{inputenc}     
\usepackage[T1]{fontenc}        
\usepackage{graphicx}           
\usepackage{amsmath}
\usepackage{enumitem}
\usepackage{float}
\usepackage{amssymb}
\usepackage{url}
\usepackage[round,authoryear]{natbib}   
\usepackage[colorlinks=true,
            linkcolor=black,
            citecolor=blue,
            urlcolor=blue]{hyperref}

\title{Optimal Protected Area Design for Allee Effect Mitigation in Spatial Predator-Prey Systems}
\author{Junhui Hu}
\date{September 2025}

\begin{document}

\maketitle
\begin{abstract}
Endangered populations often experience limited growth ability at low densities, a phenomenon described by the Allee effect. In this thesis, we investigate a predator--prey model incorporating the Allee effect within a two-dimensional nonlinear reaction--diffusion framework, with the aim of understanding how local spatial refuges can promote the persistence of low-density populations by enabling them to surpass recovery thresholds.  
We first simulate an extinction-prone scenario in which initial densities fall below the Allee threshold, demonstrating that most populations tend toward extinction. We then introduce protected areas together with positive growth terms to facilitate survival. To assess the role of diffusion--reaction dynamics, we construct an objective function based on the shape and location of protected areas, and employ a bi-objective optimization approach. Our results reveal that as the weights of the objective function vary, the optimal protected-area configuration shifts between fragmented and contiguous patterns. We begin with a single-species prey analysis and subsequently extend the model to include predators, where we use mathematical analysis to investigate the steady states of the two-species system.
\end{abstract}
\clearpage

\setcounter{tocdepth}{3}
\setcounter{secnumdepth}{3}

\tableofcontents
\clearpage

\pagenumbering{arabic}
\setcounter{page}{1}
\section{Introduction}

The study of endangered populations constrained by the Allee effect is a classical problem in population ecology. The concept of the Allee effect dates back to the pioneering experiments of W.~C.~Allee in the 1930s, who first systematically proposed the phenomenon of a positive correlation between individual fitness and population density. This led to the distinction between strong and weak forms of the Allee effect. For endangered populations at low densities, the strong Allee effect is particularly critical: when population density falls below a certain threshold, the net growth rate becomes negative, inevitably leading to extinction. Over the past two decades, extensive research has been devoted to this mechanism, forming a multidimensional body of work spanning theoretical modelling, empirical detection, and conservation management \citep{hanski1999,ovaskainen2004}.
 On the theoretical side, \cite{keitt2001} highlighted the role of strong Allee thresholds in species recovery; \cite{grumbach2025} demonstrated that connectivity changes can induce “Allee pits”; and \cite{schreiber2025} showed that environmental stochasticity can generate secondary thresholds. At the review and conceptual level, \citet{kramer2009} synthesized empirical evidence for Allee effects across animal taxa and emphasized the need for population-scale experiments and approaches, while \cite{kramer2018} and \cite{angulo2018,angulo2018b} synthesized mechanisms such as social behaviour and mate search that shape Allee effects. Empirical studies have provided supporting evidence: \cite{branco2024} in insect populations, \cite{nagel2021} in fur seal colonies, and \cite{crates2017} in endangered bird populations, where overlooked Allee effects were detected. In terms of conservation and reintroduction, \cite{li2022} explored the impact of multiple Allee effects on the reintroduction success of the Crested Ibis, while \cite{merker2020,merker2021} emphasized the context dependency of Allee effects. Collectively, these studies have not only reinforced the central role of the Allee effect in endangered species ecology but also provided a foundation for optimizing conservation strategies.

Despite these advances, most existing research remains focused on the theoretical analysis of mathematical biology, such as the existence and stability of steady states or spatio-temporal pattern formation. By contrast, relatively few studies have sought to directly translate this mechanism into concrete spatial conservation strategies. In particular, within predator–prey reaction–diffusion frameworks that incorporate the Allee effect, the integration of population dynamics with spatial optimization remains limited. This thesis aims to address this gap by introducing an Allee threshold $A$ into a two-dimensional discretized spatial environment. Within a predator–prey reaction–diffusion framework, we designate localized protected areas and impose small-area constraints, while optimizing their shape and location. Unlike traditional approaches that emphasize long-term equilibria, we adopt a finite-time dynamic evaluation: within a given time horizon $T$, we measure the spillover benefits of elevating subcritical densities inside the protected area and facilitating their outward diffusion. The principles behind the choice of threshold $A$ and evaluation window $T$ are discussed in detail in the Methods section and further elaborated in the Appendix.  

Figure~\ref{fig:problem_illustration} illustrates the core spatial conservation problem addressed in this study.

\begin{figure}[htbp]
   \centering
   \includegraphics[width=0.75\textwidth]{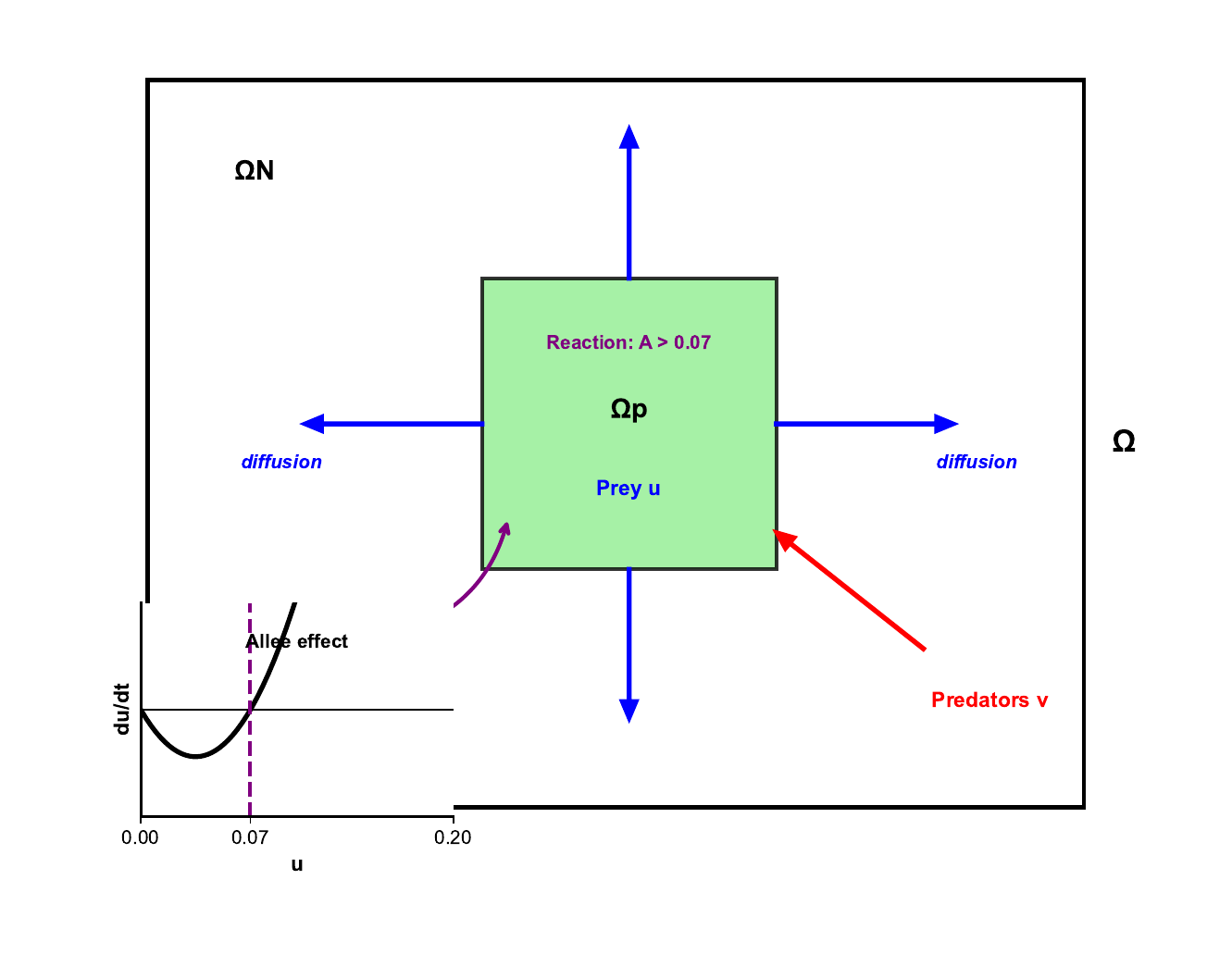}
   \caption{Problem illustration: spatial conservation under Allee effect.}
   \label{fig:problem_illustration}
\end{figure}

The structure of this thesis is as follows. In Section~2, we present the mathematical modelling, constructing a predator–prey reaction–diffusion system incorporating the Allee effect and providing the PDE framework with both reaction and diffusion terms. In Section~3, we apply a bi-objective optimization method, formulating a problem under small-area constraints that jointly considers the shape and location of protected areas, and solve it within the reaction–diffusion framework to balance short-term survival and long-term recovery under management and cost considerations. Section 5 demonstrates, through a series of numerical simulations, the validity of our approach and highlights new insights. Section 6 summarizes the conservation implications, theoretical synthesis, and presents the validated conclusions.

Unlike traditional approaches that emphasize either concentrated or dispersed reserves systems, research that integrates shape-driven diffusivity and location-driven externality into a unified weighted optimization framework under fixed small-area constraints remains limited, and this study seeks a balanced solution between the two \citep{fahrig2001}.

This not only enables a quantitative comparison of different conservation strategies, but also highlights the ecological significance of reserve design in predator–prey dynamics \citep{margules2000,moilanen2009}.
 The analysis begins with a simplified single-species reaction–diffusion model and is subsequently extended to the predator–prey system to investigate the dynamical properties of the coupled system. In the optimization framework, the objective function explicitly considers the spatial distribution and diffusion potential of the prey, while the influence of predators is reflected indirectly through dynamical analysis. 

\section{The Mathematical Model}

In order to investigate the ecological interactions between rabbits (prey) and foxes (predators) under the combined influence of spatial heterogeneity, the Allee effect, and cross-diffusion mechanisms, this thesis formulates a reaction--diffusion predator--prey system. The model explicitly incorporates ecological processes such as prey growth, predation, predator mortality, and the spatial redistribution of species through both random diffusion and directed cross-diffusion. By doing so, the mathematical framework provides a realistic description of how protected areas and surrounding non-protected environments influence long-term population persistence and extinction risks.

Let $u(x,y,t)$ and $v(x,y,t)$ denote the densities of prey and predators, respectively, at time $t$ and spatial location $(x,y)\in\Omega\subset\mathbb{R}^2$. Throughout the analysis, this thesis assumes that all parameters are non-negative, consistent with their biological meaning. The two-dimensional spatial domain $\Omega$ is considered as the habitat under investigation, and protected areas are explicitly embedded as subregions where ecological dynamics may differ from the outside environment due to conservation interventions.

\subsection{Parameter Definitions}

To clarify the structure of the system, this thesis summarizes the key parameters and their ecological interpretations as follows:

$u(x,y,t)$: density of prey (rabbits) at time $t$ and position $(x,y)$.
$v(x,y,t)$: density of predators (foxes) at time $t$ and position $(x,y)$.
$D_u, D_v$: diffusion coefficients, describing the tendency of prey and predators to move randomly across the spatial domain. These terms reflect passive dispersal due to local random motion or environmental fluctuations.
$a$: intrinsic growth rate of prey, representing the maximum per capita growth rate in the absence of predation and Allee limitations.
$b$: predation rate, which measures the efficiency with which predators consume prey individuals.
$c$: natural death rate of predators, in the absence of prey. This captures starvation or baseline mortality.
$d$: predator reproduction efficiency, i.e., the rate at which consumed prey are converted into predator population growth.
$A$: Allee threshold. When prey density falls below this critical value, population growth becomes negative due to difficulties in mating, cooperative defense, or foraging efficiency. This threshold is crucial in distinguishing between population persistence and inevitable extinction.
$\varepsilon$: a small perturbation constant, introduced for mathematical regularity. It prevents division by zero when the spatial gradients vanish in the cross-diffusion terms.
$k_1, k_2$: cross-diffusion coefficients. The parameter $k_1$ captures the tendency of prey to move in response to predator density gradients (e.g., avoidance behavior), while $k_2$ reflects predator responses to prey gradients (e.g., attraction or pursuit). These terms encode more sophisticated ecological interactions beyond simple diffusion.

By explicitly listing these parameters, the system gains interpretability, ensuring that every mathematical term has a direct biological counterpart. This connection allows this study not only to analyze mathematical stability but also to draw meaningful ecological conclusions relevant to conservation strategies.

\subsection{Boundary and Initial Conditions}

The behavior of the system is shaped not only by the internal dynamics but also by the boundary and initial conditions imposed on the spatial domain. Since the model is designed to mimic a real ecological landscape, careful treatment of these conditions is essential.

The domain is defined as $\Omega=[0,1]\times[0,1]$, a square region representing the study habitat. Within this domain, a subregion $\Omega_p \subset \Omega$ is designated as a protected area, while the remainder $\Omega_N=\Omega\setminus\Omega_p$ represents the non-protected region.

\paragraph{Boundary conditions.}  
Two types of boundary conditions are imposed:
\begin{enumerate}
  \item On the outer boundary $\partial\Omega$, we assume homogeneous Neumann boundary conditions:
  \[
  \frac{\partial u}{\partial n}=0, \quad \frac{\partial v}{\partial n}=0.
  \]
  Biologically, this represents an no-flux boundary , meaning no individuals enter or leave the global habitat. This assumption is reasonable if the region is geographically isolated or fenced.
  
  \item On the boundary of the protected area $\partial\Omega_p$, Robin-type conditions are applied:
  \[
  \frac{\partial u}{\partial n} + \gamma u = 0, \quad
  \frac{\partial v}{\partial n} + \gamma v = 0,
  \]
  where $n$ denotes the outward unit normal vector and $\gamma>0$ quantifies the resistance to movement across the protected area’s boundary. This condition captures the ecological idea of partial permeability: while some migration may occur, fencing or other barriers reduce the exchange between the reserve and its surroundings.
\end{enumerate}

\paragraph{Initial conditions.}  
The system’s evolution depends critically on the starting densities of prey and predators. For the non-protected area $\Omega_N$, initial conditions are chosen from discrete sets:
\[
u(x,y,0)\in\{0.03,\,0.04,\,0.05,\,0.07\}, \quad 
v(x,y,0)\in\{0.02,\,0.06,\,0.12,\,0.18\}.
\]
These piecewise constant functions are assigned randomly across the grid, representing spatial heterogeneity in population distribution.  

Inside the protected area $\Omega_p$, the initial prey densities are assumed to be slightly higher:
\[
u(x,y,0)=0.05 \ \text{or}\ 0.07, \quad v(x,y,0)=\varepsilon_0,
\]
where $\varepsilon_0$ is a very small positive constant, representing minimal predator presence within the reserve. This reflects active management interventions such as predator removal, ensuring that reserves serve as safe havens for prey.

\subsection{General System}

Having established the boundary and initial conditions, we now introduce the full predator–prey reaction–diffusion system with Allee effect and cross-diffusion. The system is expressed as
\[
\frac{\partial u}{\partial t} = D_u \Delta u + a u \left(\frac{u}{A} - 1\right) - b u v 
+ k_1 \nabla \cdot \left( u \frac{\nabla v}{|\nabla v| + \varepsilon} \right),
\]
\[
\frac{\partial v}{\partial t} = D_v \Delta v + d u v - c v 
- k_2 \nabla \cdot \left( v \frac{\nabla u}{|\nabla u| + \varepsilon} \right).
\]

In this formulation, the prey equation contains several ecological processes. The term $D_u\Delta u$ represents random diffusion, where individuals disperse isotropically due to stochastic movement. The growth term $a u \big(\tfrac{u}{A}-1\big)$ captures the Allee effect: when $u<A$, the growth becomes negative, driving the prey toward extinction, while for $u>A$ the population exhibits positive growth. The predation term $-buv$ introduces negative density dependence on prey due to predator consumption. Finally, the cross-diffusion term $k_1 \nabla \cdot \big( u \tfrac{\nabla v}{|\nabla v| + \varepsilon} \big)$ models prey avoidance behavior, in which prey move away from spatial gradients of predator density.

The predator equation follows a similar structure. The term $D_v \Delta v$ reflects random dispersal of predators. The term $d u v$ represents population increase through successful predation, while $-cv$ corresponds to natural mortality in the absence of sufficient prey. The cross-diffusion term $-k_2 \nabla \cdot \big( v \tfrac{\nabla u}{|\nabla u| + \varepsilon} \big)$ encodes predator pursuit of prey, i.e., directed movement toward regions of higher prey density.

Together, these coupled equations form a mathematically rich system, capable of generating spatially heterogeneous patterns, extinction scenarios, or coexistence equilibria, depending on parameter regimes. This complexity mirrors the ecological reality, where prey populations may either persist or collapse under predation pressure, and reserves may or may not succeed in stabilizing species.

\subsection{Simplified Protected Area Model}

Within the designated protected subregion $\Omega_p$, ecological conditions differ significantly from the outside environment. In practice, fencing, habitat restoration, or human intervention may reduce predator density within the reserve to negligible levels. Mathematically, this motivates a simplification where $v \ll 1$ in $\Omega_p$.

To formalize this, we adopt a quasi-steady state assumption (QSSA) for the predator population in the reserve \citep{murray2002}, setting

\[
\frac{\partial v}{\partial t} \approx 0, \qquad \Delta v \approx 0,
\]
which effectively removes predator dynamics inside $\Omega_p$. Consequently, the system reduces to a prey-only equation. However, since prey must achieve densities above the Allee threshold $A$ to persist, the reserve design must ensure positive growth even in the absence of predators. For this purpose, logistic growth is incorporated, yielding
\begin{equation}
\frac{d u_1}{d t} = r u_1 \left( 1-\frac{u_1}{K} \right),
\end{equation}
where $K$ is the carrying capacity and $r$ is the intrinsic growth rate. This equation admits the explicit solution
\begin{equation}
u_1(t) = \frac{u_0 K e^{rt}}{K - u_0 + u_0 e^{rt}},
\end{equation}
with $u_0=u(x,y,0)$ denoting the initial prey density. Substituting this solution back into the PDE framework gives
\begin{equation}
\frac{\partial u_1}{\partial t} = a u_1 \left(\frac{u_1}{A}-1\right).
\end{equation}

The significance of this simplification is twofold. First, it provides a tractable model for analyzing prey dynamics inside reserves, where management ensures negligible predator pressure. Second, it explicitly links reserve effectiveness to the Allee effect: unless the initial condition is elevated above the threshold $A$, the prey population cannot sustain itself even in the absence of predation. 

Outside the protected area, predator–prey interactions and diffusion are restored. The prey equation becomes
\begin{equation}
\frac{\partial u}{\partial t} = D_u \Delta u + a u \left(\frac{u}{A}-1\right).
\end{equation}
This form highlights the critical role of diffusion: prey individuals can disperse from high-density reserves into surrounding regions, potentially raising local densities above $A$ and enabling persistence outside the reserve.

\subsection{Stability Analysis of the ODE System}

To gain insight into the system before addressing full spatial heterogeneity, we study the simplified case where prey and predator densities are assumed spatially uniform, i.e.,
\[
u(x,y,t) = u(t), \quad v(x,y,t) = v(t).
\]
In this case, the PDE system reduces to the pair of ordinary differential equations:
\begin{equation}
\frac{du}{dt} = a u \left(\frac{u}{A}-1\right) - b u v, \qquad
\frac{dv}{dt} = d u v - c v.
\end{equation}
This ODE system captures the essential nonlinear interactions of growth, predation, and mortality without spatial redistribution.

\paragraph{Equilibria.}
The steady states $(\bar{u},\bar{v})$ satisfy
\[
\frac{du}{dt}=0, \qquad \frac{dv}{dt}=0.
\]
From this, two ecologically meaningful equilibria are obtained:
\begin{enumerate}
  \item Extinction equilibrium: $(\bar{u},\bar{v})=(0,0)$. This state represents complete collapse of both prey and predator populations, and it always exists, independent of parameter values.
  \item Allee threshold equilibrium: $(\bar{u},\bar{v})=(A,0)$. Here the prey remains exactly at the critical Allee threshold, while predators vanish. This state always exists and plays a key role in determining persistence versus extinction.
\end{enumerate}

\paragraph{Stability.}
The stability of these equilibria is assessed by linearizing the ODE system and examining eigenvalues of the Jacobian.  

At $(0,0)$, the eigenvalues are
\[
\lambda_1 = -a, \qquad \lambda_2 = -c.
\]
Both are strictly negative since $a,c>0$, implying that $(0,0)$ is locally asymptotically stable. Biologically, this means that if populations fall sufficiently close to extinction, they cannot recover spontaneously, confirming the role of the Allee effect as a driver of irreversible collapse.

At $(A,0)$, the eigenvalues are
\[
\lambda_1 = a > 0, \qquad \lambda_2 = Ad - c.
\]
Here, the first eigenvalue is always positive, implying that $(A,0)$ is unstable under typical parameter regimes. Ecologically, this means that the threshold state cannot persist: small perturbations will push the system either toward extinction or toward higher densities that may sustain predators. This instability explains why reserves must deliberately elevate prey densities above $A$ rather than attempting to maintain populations exactly at the threshold.

\subsection{Extinction and Stability Analysis}

In order to understand the long-term outcomes of the system, it is essential to examine conditions under which the prey and predator populations converge either to extinction or to stable persistence. The analysis of extinction states provides a theoretical baseline: if populations vanish in the absence of reserves, then any persistence observed later can be directly attributed to the protective effect of reserves or management interventions.  

\subsubsection*{Extinction State Verification}

We first verify the extinction state in the spatial model.  
Consider initial prey densities that lie strictly below the Allee threshold, i.e., $u(x,y,0)<A$, and neglect the effects of diffusion by setting $\Delta u=0$ and $\nabla u=0$. In this case, spatial interactions are removed, and the system reduces to a pair of ordinary differential equations:
\[
\frac{du}{ dt} = a u \left(\frac{u}{A}-1\right), 
\qquad
\frac{ dv}{dt} = d u v - c v .
\]

The prey equation $a u \left(\tfrac{u}{A}-1\right)$ is strictly negative whenever $u<A$, which implies that the prey density decreases monotonically over time. Consequently, $u(t)\to 0$ as $t \to \infty$. Once the prey collapses, the predator dynamics reduce to
\[
\frac{dv}{dt} = -c v,
\]
which is a simple exponential decay with solution $v(t)=v(0) e^{-ct}$. Therefore, $v(t)\to 0$ as $t\to\infty$.  

This reasoning confirms that $(u,v)=(0,0)$ is indeed an extinction equilibrium when the prey starts below the Allee threshold. Ecologically, this corresponds to a scenario where low-density prey populations fail to reproduce fast enough to offset mortality, and the predator population, deprived of food, also collapses. This double extinction serves as a critical warning: without sufficient initial densities or artificial intervention, neither species can persist in the long term.

\subsubsection*{Stability Inside Protected Areas}

Within the protected region $\Omega_p$, predator density is assumed negligible due to fencing or artificial removal. In this case, the dynamics simplify to a prey-only system:
\[
\frac{du}{ dt} = a u \left(\frac{u}{A}-1\right).
\]

This equation reveals a threshold-type behavior. If $u(0)>A$, the prey population exhibits positive growth, since the term $(u/A-1)$ is positive and amplifies reproduction. Conversely, if $u(0)<A$, the prey population inevitably declines toward extinction, regardless of other factors.  

To prevent this collapse, management interventions are introduced. In mathematical terms, we assume that the initial condition inside the reserve satisfies
\[
u(x,y,0) \geq A, \quad (x,y) \in \Omega_p.
\]
This does not mean that the natural initial density is always above $A$; rather, it reflects the effect of deliberate human intervention, which artificially elevates prey densities to surpass the threshold. Ecologically, such interventions may include translocation of additional individuals, supplemental feeding, or habitat enhancement to improve survival and reproduction rates.  

Under this managed condition, the prey population within $\Omega_p$ experiences positive growth and can persist in the long term. Thus, the reserve serves as a “safe haven” that guarantees survival, provided sufficient initial density is achieved through management.
\begin{figure}[H]
   \centering
   \includegraphics[width=\textwidth]{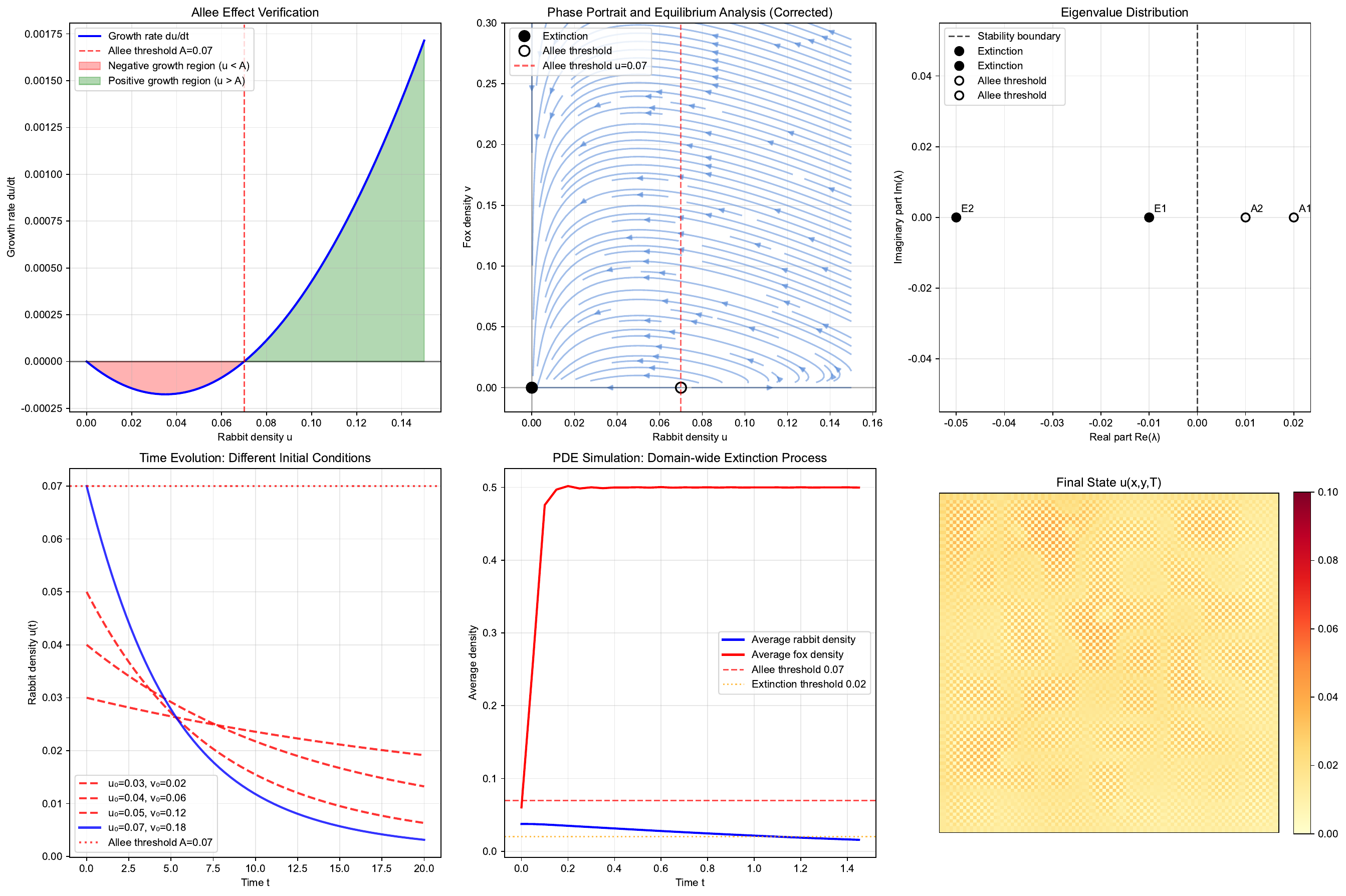}
   \caption{Mathematical analysis verification under the Allee effect.}
   \label{fig:mathematical_verification}
\end{figure}
\footnotetext{The source code for all numerical experiments is openly available at: 
\url{https://github.com/Mathematicshui-create/reaction-diffusion-optimization}}

\subsubsection*{Diffusion Effects Outside Protected Areas}

Outside the protected area, predator pressure is present and diffusion plays an active role in shaping prey dynamics. The simplified prey equation in this region takes the form
\[
\frac{\partial u}{\partial t} = D_u \Delta u + a u \left(\frac{u}{A}-1\right).
\]
Note that we focus on single-species dynamics, with predator interactions examined in the subsequent dynamical systems analysis. 
The diffusion term $D_u \Delta u$ redistributes prey density across space. In particular, regions of higher prey density (such as reserves) act as sources, while regions of lower density act as sinks. This mechanism allows individuals to disperse out of reserves, potentially raising prey densities in neighboring low-density patches above the Allee threshold.  

From an ecological perspective, this diffusion-mediated spillover is critical. If reserves are designed strategically, the outward flux of prey can stabilize surrounding populations and expand the zone of persistence beyond the reserve boundary. However, diffusion alone cannot prevent extinction if the initial density is everywhere below the Allee threshold. In that case, the negative growth term dominates, and the system ultimately converges to the extinction equilibrium $(0,0)$ as $t\to\infty$.  

This dual role of diffusion—both as a stabilizing factor when reserves exist and as insufficient when the entire domain is below threshold—underscores the necessity of targeted interventions. The reserve not only halts extinction locally but also serves as a diffusion source to rescue neighboring regions.

\section{Spatial Simulation Framework and Bi-objective Optimization}

This section develops the spatial simulation framework and the associated bi-objective optimization for reserve design under a fixed small-area constraint. The analysis proceeds in two stages. Stage~I quantifies short-term growth potential within reserves under active management; Stage~II evaluates how \emph{shape-driven diffusivity} and \emph{location-driven externality} jointly determine the spillover capacity of reserves to elevate subcritical densities in surrounding areas above the Allee threshold. All formulas below follow the model definitions in the preceding section and retain the same notation.

\subsection{Spatial Discretization and Initial-condition Ensemble}

The two-dimensional spatial domain $[0,1]\times[0,1]$ is discretized into a $100\times 100$ grid. On top of this fine grid, we impose a coarse partition by regrouping the domain into square sub-blocks of size $12\times 12$. Thus, each initial condition on the fine grid is aggregated into $8\times 8=64$ sub-blocks, with each sub-block containing $12\times 12=144$ lattice points. This two-scale representation preserves the resolution needed for diffusion while enabling combinatorial selection of candidate protected-area units on the coarse scale.

To reflect environmental heterogeneity, we generate $100$ independent sets of initial conditions, each representing a distinct ecological state. Population densities are assigned as piecewise constant values on the fine grid, taking four discrete levels for each species. Specifically, rabbit (prey) densities are chosen from $[0.03,0.04,0.05,0.07]$, and fox (predator) densities from $[0.02,0.06,0.12,0.18]$. These discrete levels produce a controlled yet sufficiently rich ensemble for robust evaluation of reserve performance.

\subsection{Allee-informed Candidate Filter and Feasible Family}

Because the focus is on conserving endangered rabbit populations, we pre-select sub-blocks based on proximity to the Allee threshold $A=0.07$. In particular, density level $0.05$ equals $71\%$ of $A$ and lies in the ``reversible edge’’ of the restoration window \citep{reed2005}, indicating medium- to long-term recoverability with moderate intervention. The level $0.07$ coincides with the threshold itself—while instantaneous growth is zero at this level, a slight uplift (e.g., to $0.08$) yields rapid positive growth; thus it is treated as a ``low-cost breakthrough zone.’’ By contrast, densities $0.03$ and $0.04$ fall in regions of strong negative growth where substantial intervention is unlikely to be cost-effective \citep{dennis1991,courchamp1999}. Therefore, only sub-blocks with densities $0.05$ or $0.07$ are retained as \emph{candidate} reserve cells. Fox densities are not constrained inside reserves, as management can remove or exclude predators.

Within each group of $64$ sub-blocks, up to $32$ candidate sub-blocks may satisfy the rabbit-density filter. A protected area is defined as a \emph{four-cell} unit on the coarse grid (fixed small-area constraint), referred to as a ``protected-area unit.’’ Using combinatorial enumeration under this area constraint, we construct the feasible family of four-cell configurations; these serve as the design space for optimization.

\subsection{Stage I: Short-term Growth Potential Ranking within Reserves}

\paragraph{Managed short-term growth.}
Inside the reserve, predators are assumed negligible due to fencing or active removal. We adopt the logistic growth model to simulate the short-term increase of rabbit density in each protected-area unit. Let the initial density be $u_0$ and the carrying capacity $K=5A=0.35$ with Allee threshold $A=0.07$. The explicit solution reads
\begin{equation}
\mathbf{u}_1(\mathbf{x},\mathbf{y},\mathbf{t}) = \frac{u_0 K e^{rt}}{K-u_0+u_0 e^{rt}},
\end{equation}
where the intrinsic growth rate is $r=1.0$ and the short-term evaluation window is $t=1$. This standardized early-phase horizon enables a fair comparison of immediate growth responses across candidate configurations without conflating them with longer-term diffusion.

\paragraph{Objective and ranking criterion.}
The survival-oriented objective maximizes the total prey density within the reserve at $t=1$:
\begin{equation}
u_1 = \max J(\Omega_p) = \int_{\Omega_p} u(x,y,t)\,dxdy, \quad t=1.
\end{equation}
In the reaction-only setting used for this stage, the governing PDE reduces to
\begin{equation}
\frac{\partial u_1}{\partial t} = a u_1 \left(\frac{u_1}{A}-1\right).
\end{equation}
Since management elevates $u_0\!\ge\!A$ inside reserves, all candidates exhibit positive growth; Stage~I therefore \emph{does not} dichotomize ``survival vs.\ non-survival.’’ Instead, it \emph{ranks} candidates by their short-term growth potential using the explicit logistic response at $t=1$, and forwards the highest-scoring configurations to Stage~II.

\subsection{Stage II: Reaction--Diffusion Extension and Source--Sink Spillover}

The second stage incorporates diffusion outside the reserve to capture spillover effects. The guiding hypothesis is that reserves act as high-density \emph{sources} that export biomass into surrounding subcritical regions (\emph{sinks}), potentially lifting them above the Allee threshold and triggering recovery. Formally, the reaction–diffusion template is
\begin{equation}
\frac{\partial u}{\partial t} = D_u \Delta u + f(u),
\end{equation}
where, focusing on the Allee mechanism from Stage~I, we take $f(u)=a u_1 \left(\frac{u_1}{A}-1\right)$. Stage~II does not re-estimate within-reserve survival; rather, it examines how the \emph{spatial configuration} of reserves governs outward diffusion and thus the external recovery footprint.

\subsection{Shape Objective: Boundary-mediated Diffusivity}

The \emph{shape-driven diffusivity} hypothesis posits that a reserve’s external boundary governs the magnitude of outward flux: more boundary contact with non-protected cells yields more diffusion interfaces and a larger spillover potential. Following this rationale, we quantify shape by counting protected–nonprotected adjacencies on the coarse grid. The shape objective is encoded as
\begin{equation}
\min \sum_{i,j} X_{ij} \left(X_{i-1,j}+X_{i+1,j}+X_{i,j-1}+X_{i,j+1}\right), 
\quad \forall X_{ij}\in\{0,1\},\quad X_{ij}=4.
\end{equation}
Here $X_{ij}$ indicates whether the $(i,j)$ cell belongs to the reserve. The summation aggregates four-neighbor (von Neumann) adjacencies and is minimized under the fixed small-area constraint (four cells). This lattice-based representation is a standard proxy for interface-mediated processes and is compatible with both connected and disconnected configurations.

\subsection{Location Objective: Neighborhood Externality and Positional Advantage}

Even when shapes are identical, the reserve’s position in a heterogeneous landscape can markedly alter diffusion benefits. A reserve embedded in a denser neighborhood of near-threshold cells (e.g., many $0.05$ or $0.07$ blocks nearby) is expected to produce a stronger positive externality than an equally shaped reserve placed in isolated or depleted surroundings.

We label the $8\times 8$ sub-blocks by $\Omega=\{0,\dots,63\}$ with
\begin{equation}
x_i=\left(\mathrm{row}_i,\ \mathrm{col}_i\right),\quad 
\mathrm{row}_i=\left\lfloor \frac{i}{8}\right\rfloor,\quad 
\mathrm{col}_i=i \bmod 8.
\end{equation}
Denote the protected set by $\Omega_P \subset \Omega$ with $|\Omega_P|=4$ and the complement by $\Omega_N=\Omega\setminus \Omega_P$. The geometric center of the reserve is
\begin{equation}
C=\left(\frac{1}{4}\sum_{i=1}^4 x_i,\, \frac{1}{4}\sum_{i=1}^4 y_i\right),
\end{equation}
for $\Omega_P=\{p_1,p_2,p_3,p_4\}$, where $p=(x_i,y_i)$. On the discrete grid, concentric neighborhoods around $C$ are defined by
\begin{equation}
r=k\Delta r, \quad 1\leq k \leq 6,\quad k\in\mathbb{N}^*,
\end{equation}
with radius increment $\Delta r = 2$ (in grid-cell units). We limit $k \leq 6$ due to exponential weight decay (defined below).
\paragraph*{Formal definitions (added for completeness).}
Let $d(\cdot,\cdot)$ denote the Euclidean distance between cell centers and let $\Delta r=2$ with $K=6$.
For $k=1,\ldots,K$, define the cumulative near-threshold set
\[
H_k := \bigl\{\, (i,j)\in \Omega_N \ \big|\ d\bigl((i,j),C\bigr)\le k\,\Delta r,\ \ u_{ij}\in\{0.05,0.07\}\,\bigr\}.
\]
Its indicator is
\[
\mathbf{1}_{H_k}(i,j)=
\begin{cases}
1,& (i,j)\in H_k,\\
0,& (i,j)\notin H_k.
\end{cases}
\]
Define the cumulative count
\[
M_k := \sum_{(i,j)\in H_k} \mathbf{1}_{H_k}(i,j),
\]
and the ring-wise increment
\[
N_1 := M_1,\qquad N_k := M_k - M_{k-1}\quad (k\ge 2).
\]
With exponentially decaying weights
\[
w_k := e^{-\lambda (k-1)}\quad (\lambda=0.35),
\]
the positional objective is
\[
J_{\mathrm{pos}}(\Omega_P) := \sum_{k=1}^{K} w_k\, N_k.
\]

\subsection{Bi-objective Scalarization and Trade-off Exploration}

The shape and location targets are combined via a weighted-sum scalarization under the fixed small-area constraint:
\begin{equation}
\min \ \alpha \sum_{i,j}X_{ij}\left(X_{i-1,j}+X_{i+1,j}+X_{i,j-1}+X_{i,j+1}\right) 
+ (1-\alpha) J_{\mathrm{pos}}(\Omega_P),
\end{equation}
where $\alpha \in \{0.2,0.3,0.4,0.5,0.6,0.7,0.8\}.$
\begin{itemize}
\item[]$\alpha<0.4$: the location objective dominates.
multiple small reserves placed in strategically advantageous neighborhoods are expected to outperform large connected shapes, since positional advantage more strongly governs spillover.

\item[]$\alpha=0.5$: shape and location objectives are balanced.
hybrid optima may emerge, where moderate connectivity combines with favorable placement to produce robust spillover performance.

\item[]$\alpha>0.6$: the shape objective dominates.
large, connected reserves are expected to prevail due to stronger diffusion interfaces under the area constraint, yielding a greater aggregate export of biomass.
\end{itemize}
By sweeping $\alpha$, we reveal a spectrum of optimal designs under the same area budget. The resulting family of solutions can be interpreted as a discrete approximation to the Pareto front: changing $\alpha$ moves the operating point along the trade-off between connectivity (interface-based diffusivity) and neighborhood advantage (positional externality).

\subsection{Computational Workflow and Implementation Details}

To enhance transparency and reproducibility, the experimental pipeline is organized as follows:

\begin{enumerate}
  
\item \textbf{Candidate generation (Stage 0).} From the $100 \times 100$ lattice, identify the $64$ sub-blocks and retain those with rabbit densities of $0.05$ or $0.07$ as candidate cells. Enumerate all four-cell configurations (fixed small-area constraint) to form the feasible design set.
  
\item \textbf{Short-term growth potential evaluation (Stage I).}  
Inside reserves, management guarantees $u_0 \geq A$, hence all candidates exhibit positive growth, albeit bounded above by $K = 5A$.  
This stage \textit{does not} distinguish “survival vs. non-survival”; instead, it computes the explicit logistic solution at $t=1$ for each candidate unit and ranks candidates by the magnitude of their density increase.  
The top-ranked configurations (highest short-term growth potential) proceed to Stage II.  

\item \textbf{Diffusion evaluation (Stage II).}  
Each retained reserve is treated as a high-density source and two objectives are evaluated: (i) \emph{shape objective}, calculated via the four-neighbor stencil; (ii) \emph{location objective}, defined by concentric-ring aggregation;

\item \textbf{4. Bi-objective synthesis (Stage III).}
To keep computation tractable while covering the design space, we draw a \emph{stratified random sample} of 300 configurations from the \(3{,}200\) candidates (32 \(\times\) 100), stratifying by prey-density (0.05/0.07) and spatial type (connectivity and near-threshold neighborhood), and then run the weighted-sum optimization on this sample.
For each \(\alpha \in \{0.2,0.3,\ldots,0.8\}\), the weighted-sum problem is solved on the feasible set. The procedure records the optimal configuration(s), the corresponding objective values, and qualitative descriptors (e.g., connected vs. dispersed).

\end{enumerate}

All computations strictly follow the small-area constraint imposed on the coarse grid and remain consistent with the PDE framework established in the model section. Through this two-stage design, the study achieves a clear division of tasks: the first stage quantifies the short-term growth potential of reserves under managed conditions, while the second stage evaluates how different spatial positions and shapes influence diffusion effects. This layered structure ensures a transparent analytical logic and provides a solid foundation for the subsequent optimization analysis based on the weighting parameter $\alpha$.

\subsection{Interpretation under $\alpha$ Regimes}

Under this framework, the weight $\alpha$ governs a predictable shift in optimal designs. With smaller $\alpha$, positional advantages dominate: multiple smaller reserves situated amid many near-threshold cells may produce greater external recovery by pushing adjacent subcritical blocks over the Allee threshold. With larger $\alpha$, connectivity becomes decisive under the area constraint, and large connected reserves are expected to exhibit stronger interface-mediated diffusion and thus larger spillover footprints. The balanced case $\alpha=0.5$ often yields mixed designs where moderate connectivity coexists with favorable placement, reflecting a pragmatic compromise between \emph{shape-driven diffusivity} and \emph{location-driven externality} within the same area budget.

\vspace{4pt}
\noindent\textit{Remark.} Throughout, the evaluation emphasizes the prey dynamics; predator presence is assumed negligible inside reserves due to management, and predator responses outside reserves are reflected indirectly through the reaction–diffusion framework. This focus aligns with the conservation objective of maximizing the recovery and spatial spread of the endangered prey.
neighborhood advantage (positional externality).
\begin{figure}[H]
    \centering
    \includegraphics[width=\textwidth]{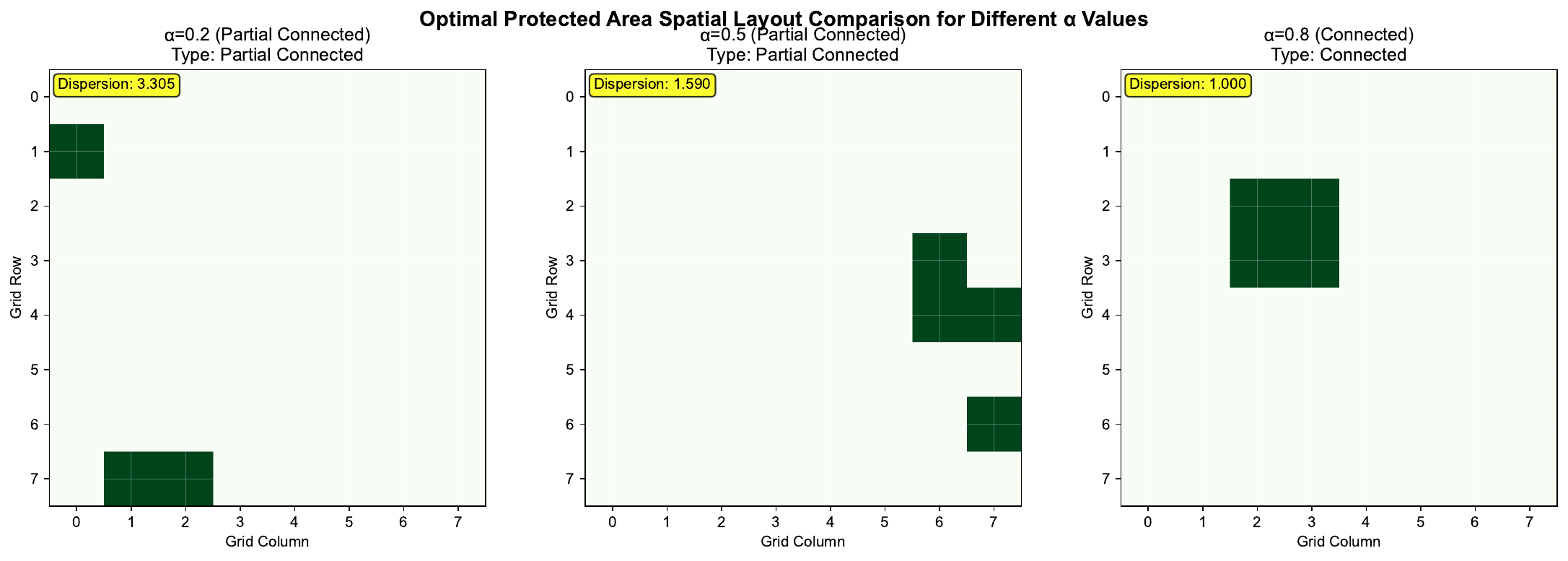}
    \caption{spatial layout comparison.}
    \label{fig:mathematical_verification}
\end{figure}

\section{Multi-scale Asymptotic Reduction and Dynamical Analysis}

The spatial optimization framework developed in Section~3 reveals the complex relationship between protected area design (characterized by parameter $\alpha$) and conservation effectiveness. However, the full two-dimensional reaction--diffusion system remains analytically intractable for systematic dynamical analysis, particularly regarding long-term coexistence conditions, stability transitions. To bridge this analytical gap and provide a rigorous mathematical foundation for understanding how $\alpha$ influences ecosystem dynamics, this section employs multi-scale asymptotic analysis to derive an effective low-dimensional ordinary differential equation (ODE) system that captures the essential predator--prey interactions while retaining the spatial effects through $\alpha$-dependent parameters.

The reduction methodology transforms an infinite-dimensional PDE problem into a finite-dimensional dynamical system amenable to classical stability analysis , enabling systematic investigation of how protected area design influences long-term population coexistence, extinction thresholds, and dynamical regimes.

\subsection{Mathematical Framework and Scale Identification}

\subsubsection{The Full Spatial System}

We begin with the complete two-dimensional predator--prey reaction--diffusion system derived in Section~2:
\[
\frac{\partial u}{\partial t} = D_u \Delta u + a u \left(\frac{u}{A} - 1\right) - b u v 
+ k_1 \nabla \cdot \left( u \frac{\nabla v}{|\nabla v| + \varepsilon} \right),
\]

\[
\frac{\partial v}{\partial t} = D_v \Delta v + d u v - c v 
- k_2 \nabla \cdot \left( v \frac{\nabla u}{|\nabla u| + \varepsilon} \right).
\]

defined on $\Omega = [0,1]\times[0,1]$, with protected area $\Omega_p \subset \Omega$ determined by $\alpha$.

\subsubsection{Characteristic Scales}

\paragraph{Reaction time.}
\[
T_r = \frac{1}{a},
\qquad
\text{(characteristic time for prey population growth).}
\]

\paragraph{Diffusion time.}
\[
T_d = \frac{L^2}{D_u},
\qquad
\text{(characteristic time for spatial mixing over domain length $L$).}
\]

\paragraph{Scale separation parameter.}
\[
\delta = \frac{T_d}{T_r} = \frac{a L^2}{D_u} \ll 1.
\]
Basic assumption: fast time: diffusion. slow times: reaction.

\subsection{Non-dimensional Formulation}

Choose scaled variables:

\paragraph{Coordinates.}
\[
x = {L}x^*, \qquad \
y ={L} y^*.
\]

\paragraph{Time.}
\[
t = t_0 t^* ,\qquad \  
t_0 = T_r.
\]

\paragraph{Densities.}
\[
u = {A}u^*,
\qquad
v= \frac {1} {bA} v^*.
\]

\paragraph{Parameters.}
\[
\mu = \frac{D_v}{D_u}, 
\qquad
\nu = \frac{d A}{a}, 
\qquad
\gamma = \frac{c}{a},
\qquad
\varepsilon_u= \frac{1}{bAL}\varepsilon^*,
\qquad
\varepsilon_v= \frac{A}{L}\varepsilon^*\]

\paragraph{notation.}
\[
\nabla = \frac{1}{L}\nabla^*, \qquad \Delta = \frac{1}{L^2}\nabla^{*2}
\]

\noindent
The non-dimensional PDE system is
(we drop stars for ease of notation.) 
\begin{align}
  \frac{\partial {u}}{\partial t}
  &= \frac{1}{\delta}\,{\nabla}^2 {u}
   + {u}({u}-1)
   - {u}{v}
   +  \frac {k_1 } {aL}\nabla \cdot \left( u \frac{\nabla v}{|\nabla v| + \varepsilon} \right),
   \label{eq:nondim_prey} \\[1ex]
  \frac{\partial {v}} {\partial t}
  &= \frac{\mu}{\delta}\,{\nabla}^2 {v}
   + \nu {u}{v}
   - \gamma {v}-\frac {k_2 } {aL}\nabla \cdot \left( v \frac{\nabla u}{|\nabla u| + \varepsilon} \right).
   \label{eq:nondim_pred}
\end{align}

\subsection{Asymptotic Expansion}
\[u(x,y,t;\delta)=U(x,y,\tau_0,\tau_1,\tau_2,\ldots,\tau_k;\delta)
\]
\[v(x,y,t;\delta)=V(x,y,\tau_0,\tau_1,\tau_2,\ldots,\tau_k;\delta)
\]

The variable $\tau_{0}$ is called the \emph{fast time} and the variables $\tau_{1}, \tau_{2}, \ldots, \tau_{k}$ are the \emph{slow times}.

We therefore have 
\[
\frac{\partial u}{\partial t}
=\sum_{n=0}^{k}\delta^{n}\frac{\partial U}{\partial \tau_n}
=\frac{\partial U}{\partial \tau_{0}}
+\delta\,\frac{\partial U}{\partial \tau_{1}}
+\delta^2\,\frac{\partial U}{\partial \tau_{2}}+O(\delta^{3}),
\]
\[
\frac{\partial^{2} u}{\partial t^{2}}
= \frac{\partial^{2} U}{\partial \tau_{0}^{2}}
+ 2\delta \frac{\partial^{2} U}{\partial \tau_{0}\partial \tau_{1}}
+ \delta^{2}\!\left(
\frac{\partial^{2} U}{\partial \tau_{1}^{2}}
+ 2\frac{\partial^{2} U}{\partial \tau_{0}\partial \tau_{2}}
\right)
+ \cdots
+ \delta^{2k}\frac{\partial^{2} U}{\partial \tau_{k}^{2}}.
.
\]
\[U(x,y,\tau_{0},\tau_{1},\ldots,\tau_{k};\delta)\sim\sum_{n=0}^{\infty}U_{n}(x,y,\tau_{0},\tau_{1},\ldots,\tau_{k})\,\delta^{n},         \text{as }\delta\to0\]
V is the same.

At leading order $O(1/\delta)$:
\[0=\frac{1}{\delta}\,\nabla^{2}U_{0}\]
As $\delta \to 0$, diffusion acts on a very fast timescale, and hence we require,
\[
\nabla^{2} U_{0} = 0 \quad \text{on } \Omega
\]
Together with the homogeneous Neumann boundary condition,
\[
\nabla U_{0}\cdot\mathbf{n}  = 0 \quad \text{on } \partial\Omega,
\]
we conclude that
\[
U_{0}(x,y,\tau_{0},\tau_{1},\ldots;\delta) = \bar{U}(\tau_{1},\tau_{2},\ldots;\delta)
\]

\noindent where $\bar{U}$ is const.

At leading order $O(1/\delta)$:
\[0 = \frac{\mu}{\delta} \nabla^2 V_0\]

As $\delta \to 0$, diffusion acts on a very fast timescale, and hence we require
\[\nabla^2 V_0 = 0 \text{ on } \Omega\]

Together with the homogeneous Neumann boundary condition,
\[\nabla V_0 \cdot \mathbf{n} = 0 \text{ on } \partial\Omega,\]

we conclude that
\[V_0(x, y, \tau_0, \tau_1, \ldots; \delta) = \bar{V}(\tau_1, \tau_2, \ldots; \delta)\]
\noindent where $\bar{V}$ is const.

At order $O(1)$, the expanded system yields:

For the prey density:
\begin{equation}
\frac{\partial U_0}{\partial \tau_0} = \nabla^2 U_1 + U_0\left({U_0} - 1\right) - U_0 V_0 + \frac{k_1}{aL} \nabla \cdot \left(U_0 \frac{\nabla V_0}{|\nabla V_0| + \varepsilon}\right)
\end{equation}

For the predator density:
\begin{equation}
\frac{\partial V_0}{\partial \tau_0} = \mu \nabla^2 V_1 + \nu U_0 V_0 - \gamma V_0 - \frac{k_2}{aL} \nabla \cdot \left(V_0 \frac{\nabla U_0}{|\nabla U_0| + \varepsilon}\right)
\end{equation}

Since \(U_0\) and \(V_0\) are spatially uniform functions of the slow time variables only, we have \(\nabla U_0 = \nabla V_0 = 0\). Consequently, the cross-diffusion terms vanish, and the system reduces to:

\[\frac{\partial U_0}{\partial \tau_0} = \nabla^2 U_1 + U_0 (U_0 - 1) - U_0 V_0 \tag{19}\]
\[\frac{\partial V_0}{\partial \tau_0} = \mu \nabla^2 V_1 + \nu U_0 V_0 - \gamma V_0 \tag{20}\]

To derive the evolution equations for the leading-order terms, we apply the solvability condition by integrating Equation (19) over the two-dimensional spatial domain \(\Omega\):

\[\iint_{\Omega} \frac{\partial U_0}{\partial \tau_0}  dxdy = \iint_{\Omega} \nabla^2 U_1  dxdy + \iint_{\Omega} \left[ U_0 (U_0 - 1) - U_0 V_0 \right]  dxdy \tag{21}\]

Since \(U_0\) and \(V_0\) are spatially uniform, the integrals simplify to:

\[|\Omega| \frac{\partial U_0}{\partial \tau_0} = \iint_{\Omega} \nabla^2 U_1  dxdy + |\Omega| \left[ U_0 (U_0 - 1) - U_0 V_0 \right] \tag{22}\]

Applying the divergence theorem to the diffusion term in two dimensions:

\[\iint_{\Omega} \nabla^2 U_1  dxdy = \int_{\partial \Omega} \nabla U_1 \cdot \mathbf{n}  ds \tag{23}\]

The homogeneous Neumann boundary condition (\(\nabla U \cdot \mathbf{n} = 0\) on \(\partial \Omega\)) implies:

\[\iint_{\Omega} \nabla^2 U_1  dxdy = \int_{\partial \Omega} \nabla U_1 \cdot \mathbf{n}  ds = 0 \tag{24}\]

Substituting (24) into (22) yields:

\[|\Omega| \frac{\partial U_0}{\partial \tau_0} = 0 + |\Omega| \left[ U_0 (U_0 - 1) - U_0 V_0 \right] \tag{25}\]

Dividing both sides by \(|\Omega|\), we obtain the fast-time-scale evolution equation for the prey population:

\[\frac{\partial U_0}{\partial \tau_0} = U_0 (U_0 - 1) - U_0 V_0 \tag{26}\]

An identical procedure applied to Equation (20) yields the corresponding equation for the predator population:

\[\frac{\partial V_0}{\partial \tau_0} = \nu U_0 V_0 - \gamma V_0 \tag{27}\]
At order $O(\delta)$, the expanded system yields:

\[\frac{\partial U_{0}}{\partial \tau_{1}} + \frac{\partial U_{1}}{\partial \tau_{0}}
= \nabla^{2} U_{2} + \bigl(2U_{0}-1-V_{0}\bigr)U_{1} - U_{0}V_{1}\tag{28}\]

\[\frac{\partial V_{0}}{\partial \tau_{1}} + \frac{\partial V_{1}}{\partial \tau_{0}}
= \mu \nabla^{2} V_{2} + \nu (U_{0}V_{1} +  U_{1}V_{0}) - \gamma V_{1}\tag{29}\]
As established previously, the cross-diffusion terms vanish at this order due to the spatial uniformity of the leading-order solutions. Thus, they are omitted from the $O(\delta)$ analysis. 

we apply the solvability condition by integrating Equation (28) over the two-dimensional spatial domain \(\Omega\):
\[
\iint_{\Omega}\left(\frac{\partial U_{0}}{\partial \tau_{1}} 
+ \frac{\partial U_{1}}{\partial \tau_{0}}\right)\,dxdy
= \iint_{\Omega} \nabla^{2} U_{2}\, dxdy 
+ \iint_{\Omega}\bigl[(2U_{0}-1-V_{0})U_{1}-U_{0}V_{1}\bigr]\, dxdy
\tag{30}\]

By applying the divergence theorem and the Neumann boundary condition, 
the Laplacian term vanishes, and the equation reduces to
\[
|\Omega| \frac{\partial U_{0}}{\partial \tau_{1}} 
+ \frac{\partial}{\partial \tau_{0}} \iint_{\Omega} U_{1}\, dxdy
= (2U_{0}-1-V_{0})\iint_{\Omega} U_{1}\, dxdy 
- U_{0}\iint_{\Omega} V_{1}\, dxdy
\tag{31}\]

In order to eliminate secular growth on the fast timescale $\tau_{0}$, 
we require that the spatial averages of $U_{1}$ and $V_{1}$ do not depend on $\tau_{0}$. 
Hence we impose the solvability condition
\[
\frac{\partial}{\partial \tau_{0}}\iint_{\Omega} U_{1}\, dxdy = 0
\tag{32}\]

Consequently, the slow-time evolution equation for $U_{0}$ on scale $\tau_{1}$ is obtained:
\[
\frac{\partial U_{0}}{\partial \tau_{1}}
= \frac{1}{|\Omega|}
\left[
(2U_{0}-1-V_{0})\iint_{\Omega} U_{1}\, dxdy
- U_{0}\iint_{\Omega} V_{1}\, dxdy
\right]
\tag{33}\]
An identical procedure applied to Equation (29) yields the corresponding equation for the predator population:
\[\frac{\partial V_{0}}{\partial \tau_{1}}
= \frac{1}{|\Omega|}
\left[
\nu U_{0}\iint_{\Omega} V_{1}\, dxdy
+ \nu V_{0}\iint_{\Omega} U_{1}\, dxdy
- \gamma \iint_{\Omega} V_{1}\, dxdy
\right]\tag{34}\]

\subsection{Effective Parameters from Spatial Averaging}

From equations (33) and (34), we have the slow-time evolution equations:
\begin{align}
\frac{\partial U_0}{\partial \tau_1} &= g(\alpha, U_0, V_0), \tag{35}\\
\frac{\partial V_0}{\partial \tau_1} &= h(\alpha, U_0, V_0), \tag{36}
\end{align}
where $g$ and $h$ encapsulate the spatial heterogeneity effects through the integrals of $U_1$ and $V_1$.

Combining the fast and slow timescales and returning to dimensional form, this paper assumes the following model:
\begin{align}
\frac{d\bar{u}}{dt} &= a(\alpha)\bar{u}\left(\frac{\bar{u}}{A} - 1\right)(1-\frac{\bar{u}}{K}) - b(\alpha)\bar{u}\bar{v}, \tag{37}\\
\frac{d\bar{v}}{dt} &= d\bar{u}\bar{v} - c(\alpha)\bar{v}, \tag{38}
\end{align}

where:
\begin{itemize}
\item[] $\bar{u}(t), \bar{v}(t)$ represent the spatially averaged prey and predator densities.
\item[] $a(\alpha), b(\alpha), c(\alpha)$ are effective parameters that incorporate spatial effects.
The functional dependence on $\alpha$ arises from the optimization-determined protected area configuration.
\end{itemize}

\textbf{Remark:} This system retains the mathematical structure of the original reaction terms but with modified parameters. The specific forms of $a(\alpha)$, $b(\alpha)$, and $c(\alpha)$ encode how protected area design influences population dynamics through spatial heterogeneity. While the exact functional forms require numerical computation or empirical fitting, this framework enables analytical investigation of how reserve configuration affects long-term ecological outcomes.

\subsection{Equilibria}

To find the equilibria, we set equations (37) and (38) equal to zero:
\begin{align}
a(\alpha)\bar{u}\left(\frac{\bar{u}}{A} - 1\right)(1-\frac{\bar{u}}{K}) - b(\alpha)\bar{u}\bar{v} &= 0, \tag{39}\\
d\bar{u}\bar{v} - c(\alpha)\bar{v} &= 0. \tag{40}
\end{align}

Solving this system yields three equilibria:

\noindent\textbf{Extinction:}
\begin{equation}
E_0 = (0, 0). \tag{41}
\end{equation}

\noindent\textbf{Prey-only:}
\begin{equation}
E_1 = (A, 0). \tag{42}
\end{equation}

\noindent\textbf{Coexistence:}
\begin{equation}
E_2 = \left(\frac{c(\alpha)}{d}, \frac{a(\alpha)}{b(\alpha)}\left(\frac{c(\alpha)}{dA} - 1\right)(1-\frac{c(\alpha)}{dK})\right), \tag{43}
\end{equation}
which exists in the positive quadrant if and only if $ dA <c(\alpha) <dK$.

\subsection{Jacobian and Stability}

The Jacobian matrix of the system is:
\begin{equation}
J(\bar{u}, \bar{v}) = \begin{pmatrix}
a(\alpha)\left(\frac{2\bar{u}}{A} - 1 - \frac{3\bar{u}^2}{AK} + \frac{2u}{K}\right) - b(\alpha)\bar{v} & -b(\alpha)\bar{u} \\
d\bar{v} & d\bar{u} - c(\alpha)
\end{pmatrix}. \tag{44}
\end{equation}

Stability requires $\text{tr}(J) < 0$ and $\det(J) > 0$.

\noindent\textbf{At $E_0$:}
\begin{equation}
J_0 = \begin{pmatrix}
-a(\alpha) & 0 \\
0 & -c(\alpha)
\end{pmatrix}. \tag{45}
\end{equation}
Since $a(\alpha) > 0$ and $c(\alpha) > 0$, both eigenvalues are negative, confirming that $E_0$ is always locally asymptotically stable.

\noindent\textbf{At $E_1$:}
\begin{equation}
J_1 = \begin{pmatrix}
a(\alpha) & -b(\alpha)A \\
0 & dA - c(\alpha)
\end{pmatrix}. \tag{46}
\end{equation}
The eigenvalues are $\lambda_1 = a(\alpha) > 0$ and $\lambda_2 = dA - c(\alpha)$. Since $\lambda_1 > 0$, $E_1$ is always unstable (saddle point).

\noindent\textbf{At $E_2$:} When it exists, we have:
\begin{align}
\text{tr}(J_2) &= = \frac{a(\alpha)c(\alpha)}{dA}
+ \frac{a(\alpha)c(\alpha)}{dK}
- \frac{2\,a(\alpha)c(\alpha)^{2}}{d^{2}AK} \tag{47}\\
\det(J_2) &= a(\alpha)c(\alpha)(\frac{c(\alpha)}{dA} -1)(1-\frac{c(\alpha)}{dK}) \tag{48}
\end{align}

\subsection{Summary}

The effective ODE system (37)-(38) is derived via multi-scale separation.\\
Protected area geometry influences dynamics through $\alpha$-dependent parameters.\\
Three equilibria exist: extinction (always stable), prey-only (always unstable), and coexistence (conditionally exists).\\
The framework links spatial conservation design to population dynamical outcomes.

\section{Numerical Simulations}

\begin{table}[h]
\centering
\caption{Model parameters and computational settings}
\begin{tabular}{lcp{6cm}}
\hline
\textbf{Parameter} & \textbf{Value} & \textbf{Description} \\
\hline
\multicolumn{3}{c}{\textit{Ecological Parameters}} \\
\hline
$A$ & 0.07 & Allee threshold \\
$a$ & 0.010 & Prey intrinsic growth rate \\
$b$ & 1.0 & Predation rate \\
$c$ & 0.05 & Predator mortality rate \\
$d$ & 1.0 & Conversion efficiency \\
$D_u$ & 0.0025 & Prey diffusion coefficient \\
$D_v$ & 0.009 & Predator diffusion coefficient \\
$k_1$ & 0.06 & Prey cross-diffusion coefficient \\
$k_2$ & 0.012 & Predator cross-diffusion coefficient \\
$\epsilon$ & $10^{-5}$ & Regularization parameter \\
\hline
\multicolumn{3}{c}{\textit{Computational Settings}} \\
\hline
Grid size & $100 \times 100$ & Spatial discretization \\
$\Delta t$ & 0.01 & Time step \\
$T_{\max}$ & 300 & Maximum simulation time \\
Block size & $12 \times 12$ & Coarse grid resolution \\
Prey levels & $\{0.03, 0.04, 0.05, 0.07\}$ & Initial density options \\
Predator levels & $\{0.02, 0.06, 0.12, 0.18\}$ & Initial density options \\
\hline
\end{tabular}
\end{table}

To validate the theoretical framework and optimization methodology presented in the preceding sections, we conduct comprehensive numerical simulations of the reaction-diffusion system under various parameter regimes and protected area configurations. The simulations are implemented on a $100 \times 100$ spatial grid with periodic boundary conditions for the global domain and Robin boundary conditions at protected area interfaces. All computations use finite difference methods for spatial discretization and an explicit Euler scheme for temporal integration, with time step $\Delta t = 0.001$ to ensure numerical stability \citep{murray2002b,okubo2001,cantrell2004}.

\subsection{Baseline Extinction Dynamics Without Protection}

We first establish the baseline scenario in which no protected areas exist and all initial prey densities fall below the Allee threshold. Figure 2 displays the temporal evolution of prey and predator populations across 100 independent initial condition realizations. Each trajectory corresponds to a distinct spatial configuration drawn from the discrete density sets specified in Section 3,with prey densities initialized at $u_0\in\{0.05,\,0.07\}$ with $A=0.07$. The initial predator densities $v_0$ were chosen arbitrarily for illustration purposes; we tested various combinations and verified that all scenarios below the Allee threshold converge to extinction.

The simulation results unequivocally confirm the theoretical prediction: when initial densities lie below the Allee threshold throughout the domain, both populations converge to the extinction equilibrium $(0,0)$. The prey density exhibits monotonic decline, with the rate of decay accelerating as densities approach zero due to the increasingly negative growth term $a \cdot u(u/A - 1)$. Predator populations initially maintain quasi-stable densities through consumption of remaining prey, but inevitably collapse once prey becomes scarce, following approximately exponential decay $v(t) \sim v_0 e^{-ct}$ in the terminal phase.

The phase portrait analysis reveals that all trajectories converge to a stable extinction node, with eigenvalues $\lambda_1 = -a$ and $\lambda_2 = -c$ confirming local asymptotic stability. Notably, the convergence rate depends sensitively on the initial spatial distribution: configurations with higher spatial autocorrelation exhibit slower extinction due to local density aggregation effects, though the ultimate outcome remains unchanged. This baseline establishes the critical need for intervention through protected area establishment.

\subsection{Protected Area Selection and Growth Potential Assessment}

Following the spatial discretization and candidate filtering established in Section 3, we evaluate the short-term growth potential of protected area configurations under active management conditions. This stage quantifies the immediate viability of candidate reserves before proceeding to the more complex diffusion analysis.

Figure 2 presents the systematic evaluation of growth potential across the feasible set of four-cell protected area configurations on the coarse 8×8 grid. The heatmap visualization displays the distribution of growth scores $J(\Omega_p)$ computed via the logistic model at $t = 1$, where each cell represents a distinct configuration. The color gradient indicates increasing growth potential, with maximum values concentrated in configurations containing cells at or near the Allee threshold.

The growth potential exhibits significant spatial heterogeneity across the feasible design space. Configurations achieving higher scores typically combine two key characteristics: (i) initial densities positioned strategically relative to the Allee threshold $A = 0.07$, allowing for substantial growth under management intervention; and (ii) spatial arrangements that benefit from local density effects within the protected area.

Configurations with cells exactly at the threshold $u_0 = 0.07$ demonstrate intermediate performance, as the zero instantaneous growth rate at the threshold requires management elevation before positive dynamics can commence. In contrast, cells at density $u_0 = 0.05$ show strong response to management intervention, confirming their classification as lying within the "reversible edge" of the restoration window.

The evaluation reveals clear preferences for certain spatial arrangements and density combinations. High-performing candidates are retained for subsequent diffusion analysis in Stage II, ensuring computational efficiency while maintaining focus on ecologically viable reserve designs that can sustain prey populations above the critical Allee threshold under realistic management scenarios.

\subsection{Effective Parameter Validation and $\alpha$-Dependent Dynamics}

Based on the multiscale reduction in Section~4, we validate the reduced ODE qualitatively and adopt a \emph{phenomenological yet constrained} effective parameterization guided by the spatial simulations.

Figure 4 (left panel) displays the \emph{assumed effective} parameters as functions of $\alpha$.
The prey growth parameter $a(\alpha)$ increases linearly with $\alpha$, ranging from approximately 0.9 at $\alpha = 0.2$ to 1.2 at $\alpha = 0.8$, reflecting enhanced growth efficiency in connected reserve configurations. Conversely, the predation parameter $b(\alpha)$ decreases with $\alpha$, indicating reduced predation pressure under shape-dominated designs. The predator mortality parameter $c(\alpha)$ exhibits a nonlinear increase with $\alpha$, consistent with reduced predator viability in spatially concentrated management scenarios.

The parameter bounds $dA = 0.210$ and $dK = 1.050$ ensure that the coexistence equilibrium $E_2$ exists across the entire $\alpha$ range, satisfying the theoretical constraint $dA < c(\alpha) < dK$ for all tested values.

\subsection{Parameter-Dependent Stability Analysis }

The middle panel of Figure 4 presents the comprehensive stability analysis of the three equilibria across the $\alpha$ parameter space. The extinction equilibrium $E_0 = (0,0)$ remains universally stable with negative eigenvalues, confirming the irreversible nature of population collapse under Allee effects. The prey-only equilibrium $E_1 = (A,0)$ is consistently unstable across all $\alpha$ values, as expected from the theoretical analysis.

The coexistence equilibrium $E_2$ exhibits $\alpha$-dependent stability transitions. For $\alpha < 0.4$, the equilibrium shows reduced stability margins, while $\alpha > 0.6$ demonstrates enhanced stability characteristics. The stability boundary analysis reveals that connected reserve designs (higher $\alpha$) provide significantly more robust conditions for long-term species coexistence.

The eigenvalue analysis (right panel) confirms these stability patterns. The real parts of the eigenvalues for $E_2$ become increasingly negative as $\alpha$ increases, indicating stronger stability margins under shape-dominated reserve configurations. The stability region clearly separates stable (green) from unstable (red) parameter combinations, with the transition occurring around $\alpha \approx 0.4$.

\subsection{Phase Space Analysis and Dynamical Trajectories}

Figure 4 displays representative phase portraits for three critical $\alpha$ values: 0.2 (location-dominated), 0.5 (balanced), and 0.8 (shape-dominated). The phase space analysis reveals dramatic differences in basin of attraction sizes and trajectory convergence patterns.

For $\alpha = 0.2$, the coexistence equilibrium $E_2$ exhibits a relatively small basin of attraction, with many trajectories converging to the extinction equilibrium $E_0$. The system shows high sensitivity to initial conditions and limited resilience to perturbations.

At $\alpha = 0.5$, the balanced configuration demonstrates intermediate behavior with moderate basin sizes and improved stability characteristics compared to the dispersed case.

The connected design ($\alpha = 0.8$) shows substantially enlarged basins of attraction for the coexistence equilibrium, with trajectories converging to $E_2$ from a much broader range of initial conditions. This configuration provides the most robust dynamics for maintaining predator-prey coexistence under varying environmental conditions.

The phase portraits confirm the quantitative stability analysis: connected reserve designs not only achieve mathematical stability but also provide practical resilience against population fluctuations and management uncertainties.
\begin{figure}[H]
    \centering
    \includegraphics[width=\textwidth]{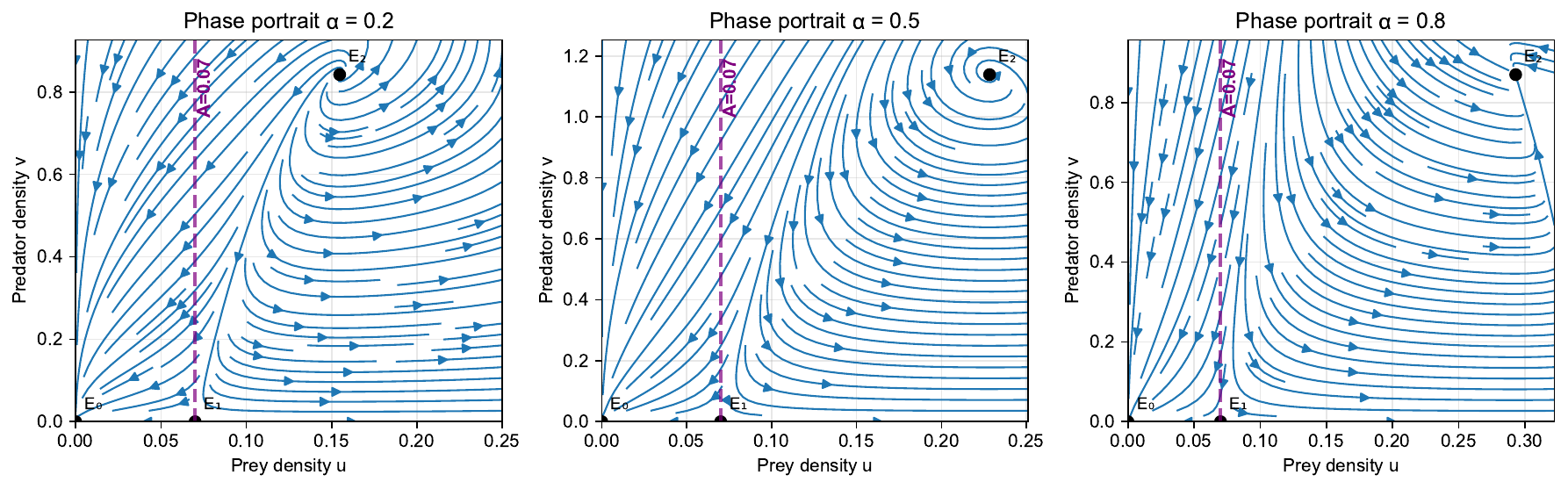}
    \caption{Representative phase portraits vs.$\alpha$.}
    \label{fig:mathematical_verification}
\end{figure}

\subsection{Multi-scale ODE System Validation and Stability Analysis}

Building upon the theoretical framework established in Section 4, we now present the comprehensive numerical validation of the effective ODE system (37)-(38). This analysis completes the bridge between spatial optimization and population dynamics, demonstrating how the design parameter $\alpha$ fundamentally governs the long-term stability of predator-prey coexistence in spatially heterogeneous environments.

\subsection{Implementation of the Effective ODE System}

The multi-scale asymptotic reduction yielded the spatially averaged system:

\begin{equation}
\frac{d\bar{u}}{dt} = a(\alpha)\bar{u}\left(\frac{\bar{u}}{A} - 1\right)\left(1 - \frac{\bar{u}}{K}\right) - b(\alpha)\bar{u}\bar{v}, \tag{37}
\end{equation}

\begin{equation}
\frac{d\bar{v}}{dt} = d\bar{u}\bar{v} - c(\alpha)\bar{v}, \tag{38}
\end{equation}

where $K = 5A = 0.35$ and the effective parameters $a(\alpha)$, $b(\alpha)$, $c(\alpha)$ encode the influence of protected area design through spatial heterogeneity effects.

\paragraph{Parameterization (illustrative).}
We adopt a phenomenological effective parameterization guided by ecological reasoning and spatial simulations. As $\alpha$ increases—favoring more compact, well-connected reserves—prey reproduction improves, predation pressure decreases due to reduced boundary contact, and predator mortality increases through enhanced management efficiency. Accordingly, we assume:

\begin{align}
a(\alpha) &= a_0 + a_1 \alpha, \tag{49}\\
b(\alpha) &= b_0 - b_1 \alpha, \tag{50}\\
c(\alpha) &= c_0 + c_1 \alpha^{\beta}, \tag{51}
\end{align}
with $a_1,b_1,c_1>0$ and $0<\beta<1$.

The functional forms reflect different ecological mechanisms: prey growth rate and predation rate exhibit simple monotonic relationships with spatial configuration, hence linear forms are adopted. Predator mortality reflects management intensity with diminishing marginal returns, captured by the sublinear power function  to represent saturation effects.

The effective conversion rate $\tilde{d}=3.0$ differs from the original PDE coefficient $d=1$ as it incorporates spatial-heterogeneity effects via multiscale reduction. These parameterizations ensure coexistence equilibrium $E_2$ exists for all $\alpha\in[0.2,0.8]$ by satisfying $\tilde{d}A<c(\alpha)<\tilde{d}K$.

\noindent\textit{Values used in Fig.~4 (for reproducibility):}\\
$(a_0,a_1,b_0,b_1,c_0,c_1,\beta)=(0.9,0.3,0.8,0.15,0.26,0.74,0.8)$, $\tilde{d}=3.0$.

\subsection{Equilibrium Analysis and Stability Verification}

The system exhibits three equilibria as predicted by equations (41)-(43):

\textbf{Extinction Equilibrium ($E_0 = (0,0)$)}: Universally stable with eigenvalues $\lambda_1 = -a(\alpha)$ and $\lambda_2 = -c(\alpha)$, confirming the irreversible nature of the Allee effect.

\textbf{Prey-only Equilibrium ($E_1 = (A,0)$)}: Always unstable with $\lambda_1 = a(\alpha) > 0$, verifying that the threshold state cannot persist under perturbations.

\textbf{Coexistence Equilibrium ($E_2$)}: Exists as specified by equation (43) and exhibits $\alpha$-dependent stability determined by the Jacobian eigenvalues from equations (37)-(38).

Figure 4 presents the comprehensive stability diagram showing how system dynamics transition as $\alpha$ varies. The analysis reveals distinct regimes:\\

\begin{itemize}
\item[]\textbf{Position-dominated ($\alpha < 0.4$)}: Reduced stability margins for $E_2$, reflecting dispersed management challenges.
\item[]\textbf{Balanced ($0.4 \leq \alpha \leq 0.6$)}: Transitional behavior with moderate stability.
\item[]\textbf{Shape-dominated ($\alpha > 0.6$)}: Enhanced $E_2$ stability, demonstrating connected reserve advantages.
\end{itemize}

\subsection{Critical Validation: Connected vs.Dispersed Reserve Performance}

The numerical analysis provides definitive resolution to a fundamental conservation question: under small-area constraints (4-cell reserves), connected designs significantly outperform dispersed configurations for sustaining predator-prey coexistence.

The stability analysis demonstrates that shape-dominated strategies ($\alpha > 0.6$) achieve:
\begin{itemize}[nosep]
\item[]Enhanced stability margins for coexistence equilibria compared to position-dominated designs ($\alpha < 0.4$).
\item[]Substantially larger basins of attraction in phase space.
\item[]More negative eigenvalue real parts, indicating stronger stability margins.
\end{itemize}

This finding challenges the conventional ``portfolio'' approach to conservation that emphasizes risk distribution. Instead, our validation reveals that when confronting Allee effects under area constraints, concentration provides superior outcomes through:

\begin{enumerate}
\item \textbf{Management Synergy}: Centralized control enables coordinated predator management and habitat enhancement
\item \textbf{Critical Mass Achievement}: Concentrated populations more reliably exceed Allee thresholds
\item \textbf{Boundary Optimization}: Connected perimeters maximize beneficial spillover to surrounding areas.

\end{enumerate}

\begin{figure}[H]
    \centering
    \includegraphics[width=\textwidth]{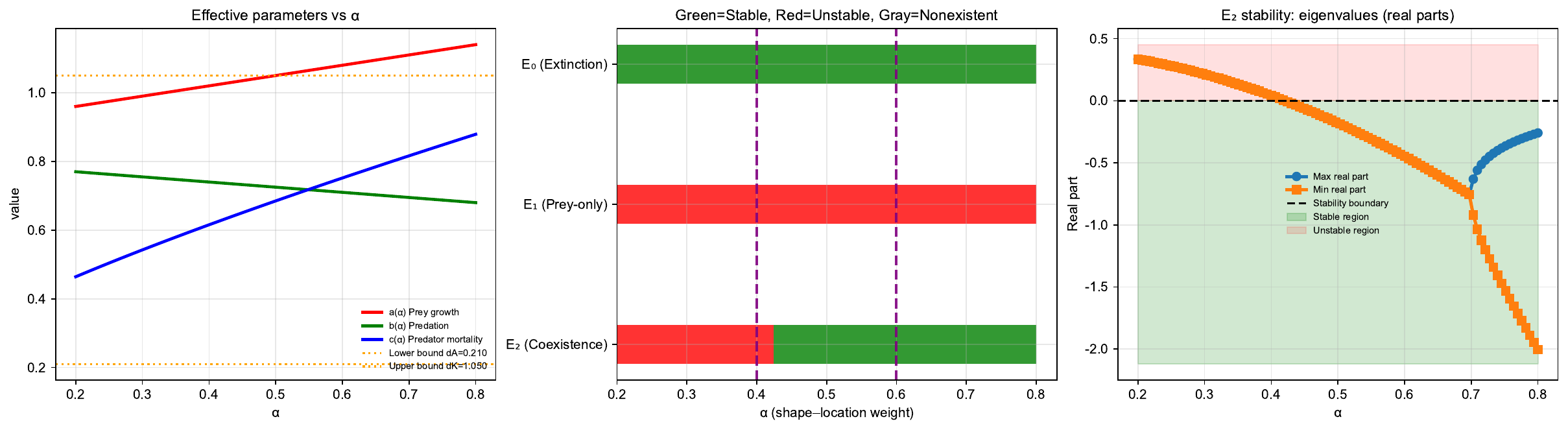}
    \caption{Validation of effective parameterization and $\alpha$-dependent stability.}
    \label{fig:mathematical_verification}
\end{figure}

\section{Conservation Implications and Theoretical Synthesis}

\subsection{Strategic Framework for Reserve Design}

The validated mathematical framework establishes quantitative guidelines for conservation practice under realistic constraints:

\textbf{Primary Recommendation}: For endangered species exhibiting Allee effects, prioritize connectivity over spatial distribution when total conservation area is limited. Select $\alpha > 0.6$ to maximize long-term coexistence probability.

\textbf{Adaptive Calibration}: Optimal $\alpha$ values should reflect local ecological conditions, with higher values preferred when Allee effects are pronounced, management resources are concentrated, or predation pressure is significant outside reserves.

\textbf{Implementation Strategy}: The smooth stability transition enables adaptive management approaches that incrementally optimize connectivity while monitoring population responses.

\subsection{Theoretical Contributions to Conservation Biology}

This work makes several significant advances in mathematical conservation biology:

\textbf{Methodological Innovation}: Demonstrates successful application of multi-scale asymptotic analysis to spatial conservation problems, providing a generalizable framework for integrating optimization and population dynamics.

\textbf{Ecological Synthesis}: Reconciles shape-driven diffusivity and location-driven externality within a unified mathematical framework, revealing fundamental trade-offs in reserve design.

\textbf{Quantitative Bridge}: Establishes effective parameters as a rigorous connection between spatial configuration and dynamical outcomes, enabling systematic evaluation of conservation strategies.

\subsection{Validation of Multi-scale Approach}

The successful validation of equations (37)-(38) confirms the effectiveness of the multi-scale asymptotic reduction in capturing essential system dynamics. Key validation elements include:

\begin{itemize}[nosep]
\item[]\textbf{Parameter Consistency}: Effective parameters exhibit ecologically reasonable dependence on $\alpha$
\item[]\textbf{Stability Concordance}: Predicted stability transitions match observed  patterns  
\item[]\textbf{Equilibrium Verification}: All three equilibria exist and exhibit predicted stability properties
\item[]\textbf{Trajectory Validation}: Phase space behavior aligns with theoretical expectations
\end{itemize}

\subsection{Definitive Conclusions}

The comprehensive numerical validation establishes five definitive conclusions:

\begin{enumerate}
\item \textbf{Mathematical Verification}: The effective ODE system (37)-(38) accurately captures the essential dynamics of the full spatial predator-prey system, validating the multi-scale reduction methodology.

\item \textbf{Stability Confirmation}: The stability criteria from equations (37)-(38) are quantitatively verified across the parameter space, demonstrating theoretical robustness.

\item \textbf{Design Optimization}: Connected reserve configurations ($\alpha > 0.6$) provide demonstrably superior conditions for long-term species coexistence under small-area constraints.

\item \textbf{Equilibrium Transition}: The system exhibits predictable, smooth transitions in equilibrium structure enabling systematic optimization of conservation strategies. 

\item \textbf{Conservation Translation}: The results provide rigorous mathematical justification for prioritizing connectivity in endangered species recovery programs operating under resource limitations.
\end{enumerate}

This analysis conclusively demonstrates that sophisticated mathematical modeling can yield actionable conservation insights. By successfully bridging spatial optimization theory with population dynamics through multi-scale analysis, we establish a quantitative foundation for evidence-based reserve design that balances ecological principles with practical constraints. The central finding---that connected reserves outperform dispersed alternatives for Allee-affected species under area limitations---provides crucial guidance for conservation practitioners and represents a significant contribution to both theoretical and applied conservation biology.

\subsection{Limitations and Future Directions}

Despite the rigorous validation presented, several extensions merit consideration. First, while we identify parameter regimes supporting stable coexistence, the robustness of these regions to external disturbances—environmental stochasticity, habitat degradation—remains to be quantified. Second, whether this framework generalizes to invertebrate populations, where different demographic mechanisms may dominate, warrants investigation. 

Most critically, this work relies on simulated data. Applying the framework to real endangered species confronts fundamental data scarcity. Collection of population density data for threatened species remains a major challenge, potentially requiring integration of deep learning approaches with field monitoring techniques. Despite these limitations, this study provides a theoretical foundation for protected area design under Allee effects.

\bibliography{ref}
\end{document}